# MEASUREMENT OF LOW ENERGY COMPONENT OF THE FLUX OF COSMIC RAYS USING NUCLEAR TRACK DETECTORS

ANA CHIRIACESCU[1,2,+], I. LAZANU[1,*]

[1] University of Bucharest, Faculty of Physics, Bucharest-Magurele, POBox MG 11, Romania,
[2] Horia Hulubei National Institute of Physics and Nuclear Engineering,
Bucharest-Magurele, 077125 Bucharest, Romania
[+] E-mail: anachiriacescu@gmail.com
[*] E-mail: ionel.lazanu@g.unibuc.ro

**Abstract** The effects induced by muons with very low energies are usually neglected. In fact, they could represent a source of radioactive background due to capture processes in different materials, which in most of cases produce radioactive isotopes, and thus they must be taken into account. Plastic track detectors have been used in the present paper to measure the ratio between the vertical and horizontal components of the flux of very low energy terrestrial muons at ground level. The data have been collected during 160 days.

## 1. Introduction

A good knowledge of all the radioactive background sources is vital for large classes of physics experiments, and in particular studies of rare processes in nuclear or particle physics, dosimetric evaluations for population or professionals, specific effects in electronic devices, as for example single error effects and soft error effects. In this context, an important source of radioactive background is the cosmic radiation also including its low energy component that usually is neglected.

The purpose of this study is to measure the angular distribution of cosmic rays, especially the low and very low energy component of muons, which can generate radioactive isotopes or particles, thus producing cosmogenic activation, and also to obtain a proper understanding of the response of plastic tracking detectors and to find the optimal detection techniques and etching conditions for them.

## 2. Interaction of muons with matter

Muons are elementary particles with one electric charge, but are about 200 times heavier than electrons, and only interact electromagnetically and weakly. The negative muon also has an antiparticle, the positive muon, both decaying spontaneously, with a mean lifetime of 2.2 μs, into one electron/positron and one neutrino and one antineutrino such that the conservation of leptonic numbers is fulfilled.

Muons can interact with matter at low and very low energies producing three major classes of processes: direct ionization, electromagnetic scattering on the nuclei and capture in the nuclei of the target [1, 2, 3].

Muons are the dominant component of charged particles at ground level and have a great penetrability in matter. The integral fluxes averaged over a solar cycle versus their kinetic energy for geomagnetic latitude around 40° (see figure 1 from reference [4]) evidence the dominant contribution of low energy muons, below 10 GeV.

In this paper the ratio between vertical and horizontal components of the flux of the low energy terrestrial muons at ground level is measured using plastic track detectors in two ways: directly in the plastic track detectors and by placing a silicon wafer above the detector, so that muons close to stop are captured in silicon, and the reaction products (charged particles) are identified in the detector.

The arguments in favor of the use of silicon are related to the good knowledge of the interaction muon - silicon from both theoretical and experimental points of view [5], [6], [7]. When negative muons are stopped into silicon, 35% of them decay into an electron and two neutrinos and the remaining 65% are captured [8]. In this case the capture process is expected to be produced by the following reaction:

$$\mu^- + {}^{28}Si \rightarrow {}^{28}Al^* + \nu + 100.5 \text{ MeV} \tag{1}$$

As an elementary process, the muon interacts with a proton from nucleus producing a neutrino and a neutron. Most of the rest energy of the muon (~100 MeV) is transferred to the neutrino and thus the energy of the neutron recoil is low. The mean excitation energy of the nucleus is approximately 15 – 20 MeV, therefore the nucleus can emit photons, one or more neutrons or charged particles in order to reach its ground state. In light nuclei, for example in silicon, the Coulomb barrier is not high and emission of multiple particles can occur. The compilation from reference [9] presented in Table 1 indicates an average of 17.5% yields for charged particle emission.

*Table 1.*
Yields of muon capture reactions (%) in silicon [9]

| Reaction | Yield (%) |
|---|---|
| (μ,ν) | 26 |
| (μ,νn) | 49 |
| (μ,ν2n) | 6 |
| (μ,ν3n) | 1 |
| (μ,ν4n) | 0 |
| (μ,νp) | 3 |
| (μ,νpn) | 9 |
| (μ,νp2n) | 2.5 |
| (μ,ναxn) | 3 |

From the same reference, the threshold energies of emission of different particles after muon capture, presented in Table 2 indicate that single particle emission is the favored process.

*Table 2.*
Threshold energy for particle emission after muon capture in $^{28}Si(\mu,x)^{28}Al$ [9]

| Emission of particle(s) | Threshold energy for emission [MeV] |
|---|---|
| n | 7.7 |
| p | 9.6 |
| d | 13.8 |
| α | 10.9 |
| 2n | 20.8 |
| 3n | 32.1 |
| 4n | 49.1 |

In the present study only charged particles with sufficient kinetic energy to produce identifiable defects in plastic detectors are measured. The excited aluminum nucleus decays by one of the following reactions:

$${}^{28}Al^* \rightarrow {}^{27}Al^* + n \text{ (12.4 MeV)} \tag{2}$$
$${}^{28}Al^* \rightarrow {}^{27}Mg + p \text{ (14.2 MeV)} \tag{3}$$
$${}^{28}Al^* \rightarrow {}^{24}Na + \alpha \text{ (15.5 MeV)} \tag{4}$$

$$^{28}Al^* \rightarrow {}^{26}Mg + d \ (18.4 \text{ MeV}) \qquad (5)$$

The energies between parentheses refer to the ground states with respect to silicon.

## 3. Experimental method, setup and results

The measurements were performed in the laboratories of the University of Bucharest, at Măgurele. The plastic track detectors TASTRAK, produced by TASL – UK [10] are polyallyl diglycol carbonate (PADC) with the chemical composition $C_{12}H_{18}O_7$. The material is clear, colourless, rigid, with density 1.30 g.cm$^{-3}$, and thicknesses of 1.0 mm. After exposure, the tracks were revealed by etching the material for 6.5 hours at 70 °C in a NaOH solution in water (concentration 6.25 N).

The detectors were placed in air, collimated by 5 cm of spectroscopic lead and shielded by 10 cm of the same material. In front of the experimental setup, at about 10 cm, a glass screen (~ 6 mm thickness) was placed, in order to absorb the charged components from cosmic rays. These detectors were exposed to terrestrial cosmic rays continuously for 160 days.

In order to investigate muon capture in Si, samples of single crystal Si, 10x10 mm$^2$ area and 300 μm thickness were placed above the detectors. Taking into account the energy dependence of the muon range in silicon, we found that muons of energy lower than 3MeV can produce interactions or capture at stop inside the target. We would like to mention that each plastic detector has a part covered by silicon, and another one directly exposed to the particle flux.

Because there are no measurements of low energy cosmic rays (in particular muons) for this location (Măgurele, Romania) except the measurements using the WILLI detector [11], a simulation program analyzing a period of 25 years was used [12]. In this code, the flux calculations were performed using the PARMA analytical model. From the time dependence of the total flux of negative muons represented in Figure 1, one can observe an oscillation of 11 years with maximum variations of around 14%.

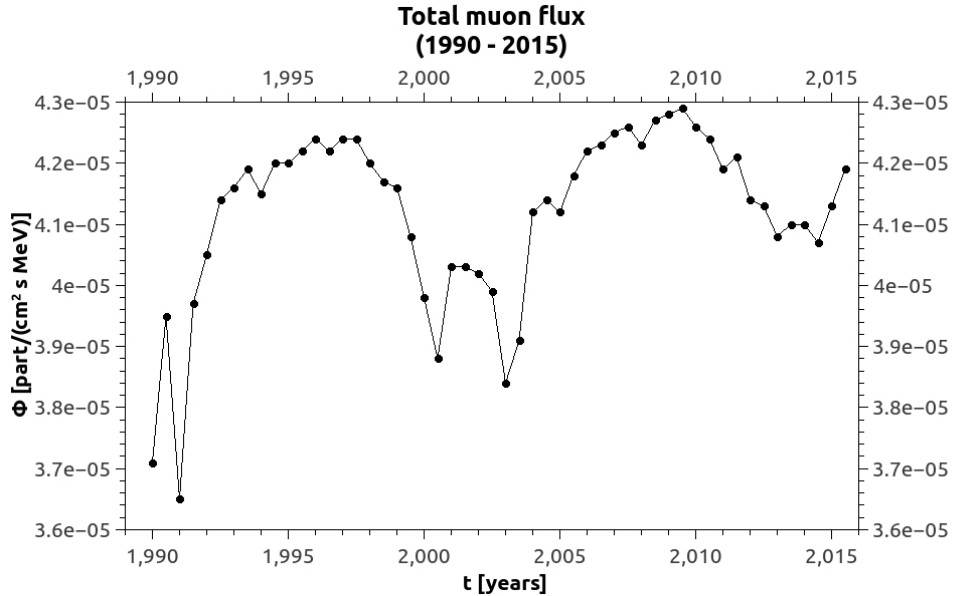

Fig.1. Total muon flux variation for 25 years in Măgurele

As the nuclear capture happens for low energy muons, it is important to know the flux variations for this energy range. For this purpose, the calculations were done for negative muons with energies below 3.0 MeV. In this case, although there is a positive slope, the variations of the flux are of the order of (0.1 – 7) x10$^{-9}$ (part/cm$^2$ s MeV). The results are presented in the Figure 2.

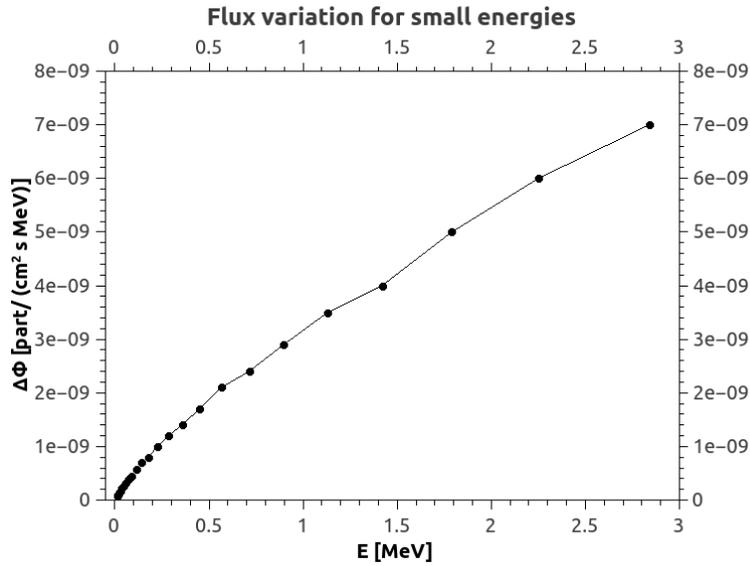

Fig. 2. Flux variation for negative muons with energies below 3.0 MeV

In order to obtain numerical results, the surface of the etched plastic detectors was scanned using a Leitz - Wetzlar optical microscope coupled with an Optika digital camera. The tracks of the charged particles produced were identified and their parameters measured.

In Figures 3a) and 3b) two examples of registered events as direct tracks are presented. Fig.3a shows events recorded at 30° inclination with respect to the vertical direction, while in Fig 3b events measured in the horizontal direction are presented.

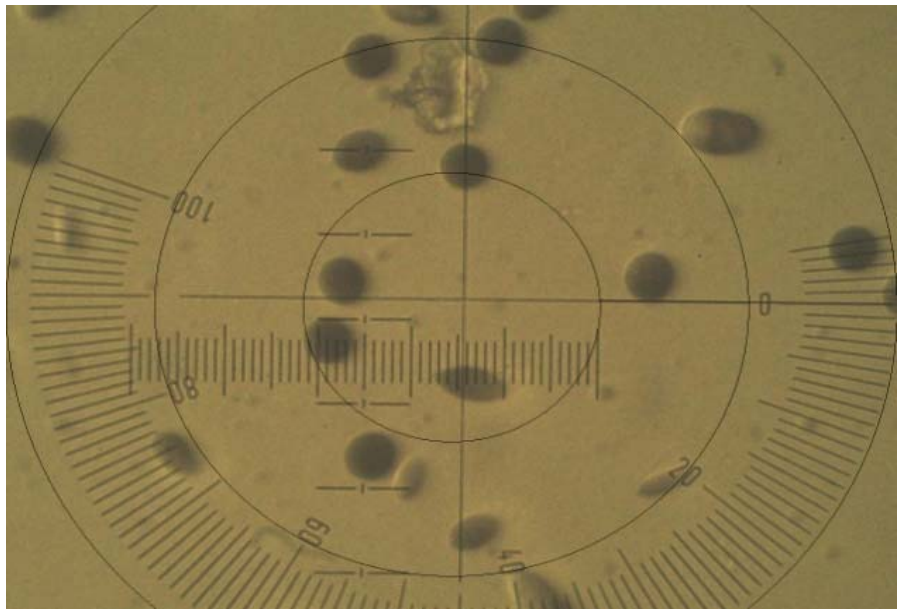

Fig. 3a. Charged particle flux – 30° in respect to vertical component

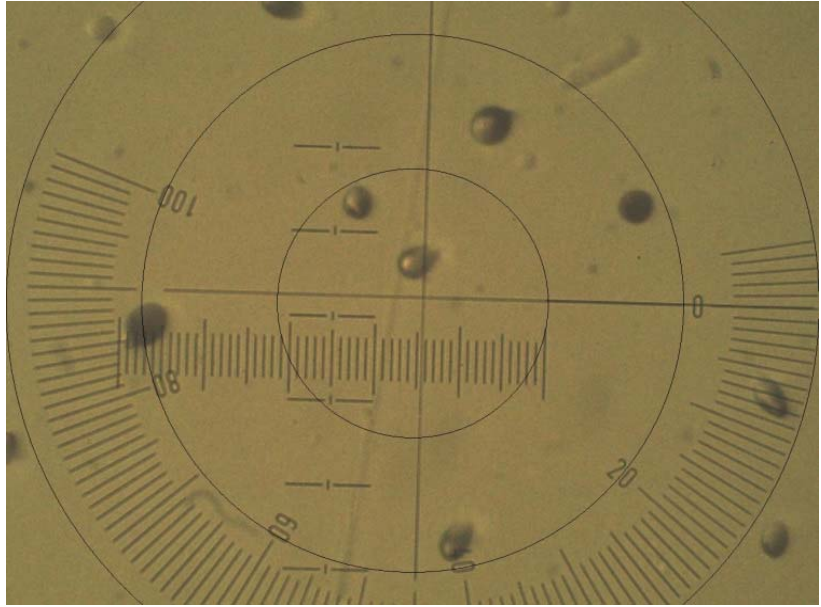

Figure 3b. Charged particle flux – horizontal component

Using the total number of particles and considering a dependence on the zenith angle of the type $I_0 \cos^2 \theta$, we obtained for the whole energies of muons able to be registered in plastic detectors a ratio between vertical and horizontal components equal to (2.8 ± 0.5). Using only charged particles produced from muon capture, which represent only 0.15±0.02 per capture as was determined in silicon by Sobottka and co-workers [8], the corresponding ratio obtained is (2.4 ± 2.1) close to previous values but with a great uncertainty.

## 4. Conclusions

This study is a preliminary investigation for a proper understanding of the response of plastic detectors TASTRACK, for finding optimal measurement and etching methods/conditions, as a necessary intermediate step before the start of a systematic study of this energy region of cosmic rays.

In the low statistics limit, the first measurements of the components of the low energy cosmic muons were performed. A ratio between vertical and horizontal components equal to (2.8 ± 0.5) was found for the direct exposure of plastic detectors, and of (2.4 ± 2.1) for the exposure through silicon single crystals, respectively. In the last case, only charged particles produced by muon capture in silicon are detected.

## Acknowledgements

This work was partially supported by the Programme CERN-RO, under Contract NEPHYLAro.